\documentclass[apj]{emulateapj}
\usepackage{apjfonts}

\slugcomment{Accepted for publication in The Astrophysical Journal}

\shorttitle{$\nu$-Process in Pop III Core-Collapse SNe}
\shortauthors{Yoshida et al.}

\begin{document}


\title{$\nu$-Process Nucleosynthesis in Population III 
Core-Collapse Supernovae}


\author{Takashi Yoshida\altaffilmark{1}, Hideyuki Umeda\altaffilmark{2},  
and Ken'ichi Nomoto\altaffilmark{2,3}}
\affil{$^1$National Astronomical Observatory of Japan, Mitaka, 
Tokyo 181-8588, Japan}
\email{takashi.yoshida@nao.ac.jp}
\affil{$^2$Department of Astronomy, School of Science, University of Tokyo,
Tokyo 113-0033, Japan}
\affil{$^3$Institute for the Physics and Mathematics of the Universe, 
University of Tokyo, Chiba 277-8582, Japan}




\begin{abstract}
We investigate the effects of neutrino-nucleus interactions 
(the $\nu$-process) on the production of iron-peak elements in Population III
core-collapse supernovae.
The $\nu$-process and the following proton and neutron capture reactions
produce odd-$Z$ iron-peak elements in complete and incomplete Si burning 
region.
This reaction sequence enhances the abundances of Sc, Mn, and Co
in the supernova ejecta.
The supernova explosion models of 15 $M_\odot$ and 25 $M_\odot$ stars 
with the $\nu$-process well reproduce the averaged Mn/Fe ratio observed 
in extremely metal-poor halo stars.
In order to reproduce the observed Mn/Fe ratio, the total neutrino energy 
in the supernovae should be $3 - 9 \times 10^{53}$ ergs.
Stronger neutrino irradiation and other production sites are
necessary to reproduce the observed Sc/Fe and Co/Fe ratios, 
although these ratios increase by the $\nu$-process.

\end{abstract}





\section{Introduction}

The abundance-metallicity relation of low mass extremely metal-poor (EMP) 
stars ($-4 \la [{\rm Fe/H}] \la -3$; $[{\rm X/Y}] \equiv \log_{10} 
(N_{\rm X}/N_{\rm Y}) - \log_{10} (N_{\rm X}/N_{\rm Y})_\odot$ 
where $N_{\rm X}$ and $N_{\rm Y}$ are the abundances of elements X and Y, 
respectively) and very metal-poor (VMP) stars 
($-3 \la [{\rm Fe/H}] \la -2$) has been clarified by observations of very 
high quality spectra \citep[e.g.,][]{nr01,ca04}.
This relation is expected to provide the information of supernovae (SNe) of 
Population (Pop) III (the first generation) massive stars 
\citep[e.g.,][]{nt06} and the first stage of Galactic chemical 
evolution (GCE) \citep[e.g.,][]{kn06}.
The EMP stars are considered to have suffered the injection of heavy 
elements from one or a few SNe \citep[e.g.,][]{st98}.
A halo interstellar medium becomes gradually homogeneous with metallicity
\citep{as00}.
Although recent studies of the star formation in the first stage of GCE
have been in progress, the characteristics of first generation stars,
such as initial mass function, have not been clarified.
In order to clarify such characteristics, we should investigate  
nucleosynthesis in the candidates of the first generation stars as well as 
the observed variations of the abundance distributions of the EMP and
VMP stars.

Recent observations have indicated that the abundance ratios [X/Fe] 
averaged in the metal-poor stars with $-4 \la [{\rm Fe/H}] \la -2$ are 
between $-0.5$ and 0.5 for most of observed elements \citep{ca04}.
Their scatter around the average values deduced from the new observation is
much smaller than found in earlier studies.
The elemental abundance distribution of the metal-poor stars is not so 
different from that of the solar-system composition.
Therefore, the small scatter is considered to suggest primordial burst
of high-mass stars or the very rapid mixing of matter from different
bursts of early star formation.

From theoretical viewpoint, \citet{tu07} compared the Pop III hypernova (HN)
yields with the abundance pattern of EMP stars with the metallicity range
$-4.2 < [{\rm Fe/H}] < -3.5$ in \citet{ca04}.
They adopted the mixing-fallback model for the HNe.
The HN yields give good agreement with the observed abundances for
C, Na, Mg, Al, Si, Ca, Ni, and Zn.
They also compared the Pop III SN and HN yields integrated
over the Salpeter initial mass function (IMF) with the abundances of VMP
stars in the metallicity range $-2.7 < [{\rm Fe/H}] < -2.0$ in \citet{ca04}.
They reproduced the observed trends of abundance ratios for
C, O, Na, Mg, Al, Si, Ca, Cr, Ni, and Zn.
The SN and HN yields of N, K, Sc, Ti, Mn, and Co are short for explaining 
the observed yield ratios, although Sc and Ti may be improved in the
high-entropy explosion models.
The yields of Mn and Co may be reproduced by the modification of 
electron fraction $Y_e$, of which value is still uncertain in the 
innermost region of SNe and HNe.
Additional nucleosynthesis processes also remain open for these elements.


During a SN explosion, the collapsed core becomes extremely high temperature 
and high density, and neutrinos are produced through pair creation.
The neutrinos are almost thermalized by neutrino-nucleus scattering
at the center.
They pass away through the neutrino sphere with carrying the gravitational
binding energy of the core.
The neutrino irradiation lasts for about $\sim 10$ s.
The total number of neutrinos emitted in a SN is very huge, i.e., 
about $N_\nu \sim 10^{58}$.
The neutrinos interact with nuclei in the surrounding stellar material 
to produce new species of nuclei.
This is called the $\nu$-process.

The $\nu$-process is expected to be 
an important production process for some elements in Pop III SN 
explosions.
The $\nu$-process is important for the synthesis of light elements,
Li and B \citep{de78,wh90,ww95,yt04,yk05,yk06a,yk06b}, 
F \citep[e.g.,][]{wh90,ww95}, and some heavy neutron-deficient nuclei 
such as $^{138}$La and $^{180}$Ta \citep{ga01,hk05}.
The $\nu$-process products are synthesized through the spallations by
neutrinos from abundant seed nuclei.
It also produces protons and neutrons and their capture reactions also 
enhance the abundances of some elements.
For Pop III SNe, the seed nuclei are $\alpha$-nuclei
abundantly produced in complete and incomplete Si burning during the
explosions.
Thus, the $\nu$-process in Pop III SNe will produce 
additional some odd-$Z$ elements.

We note that most of the nucleosynthesis calculations including the 
$\nu$-process have been carried out using spherical explosion models
\citep{ww95,yu05}.
Although the total neutrino energy should be close to the gravitational
binding energy released from the collapsing core, properties of the
neutrinos are still uncertain.
The properties strongly depend on explosion mechanism but details of 
the mechanism have not been clarified.
On the other hand, aspherical explosions have been investigated
recently \citep[e.g.][]{ks04,sp04}.
In such a case, aspherical structure of neutrino sphere can be formed
and the properties of the neutrinos are also still uncertain.
Therefore, the dependence of the abundances of Pop III SNe and
HNe on neutrino properties should be investigated.

In this study, we focus on the effect of the $\nu$-process 
on the production of odd-$Z$ iron-peak elements, Sc, Mn, and Co, during 
Pop III SN and HN explosions.
We calculate the SN nucleosynthesis including the $\nu$-process.
Then, we investigate the dependence of the abundances of Sc, Mn, and Co
in the SN ejecta on the energy of neutrinos, the stellar mass, and 
the explosion energy.
Sc, Mn, and Co are expected to be at least partly synthesized through 
the $\nu$-process.
We also indicate whether the abundance ratios of Sc, Mn, and 
Co to Fe reproduce those observed in the EMP stars 
when we take into account the $\nu$-process.

In \S 2 we explain stellar evolution models and the explosion models of
SNe and a HN.
We also describe the nucleosynthesis models including the $\nu$-process 
and the neutrino irradiation models.
In \S 3 we show the abundance distributions of Sc, Mn, and Co in the ejecta 
of the SN and HN models.
We explain additional production of these elements through the 
$\nu$-process. 
In \S 4 we explain the effect of the $\nu$-process on the abundance ratios 
of Sc, Mn, and Co to Fe and the dependence on the neutrino irradiation 
strength.
We also compare our results with the abundance ratios observed in the
EMP stars.
In \S 5 we discuss other production processes of Sc, Mn, and Co proposed 
in recent studies and the effect of the $\nu$-process.
Finally, we summarize this study in \S 6.

\section{SN Explosion Models for Population III Massive Stars}

\subsection{Stellar Evolution and SN Explosion Models}

We calculate the evolution of Population III stars for initial masses of
15 $M_\odot$ and 25 $M_\odot$ from zero-age main-sequence to the onset
of the core collapse.
We use a Henyey-type stellar evolution code including a nuclear reaction 
network with about 300 isotopes to calculate detailed nucleosynthesis and 
energy generation.
The input physics is the same as the models used in \citet{un02}.
SN explosions are calculated using a hydrodynamical code of 
spherically symmetrical piecewise parabolic method \citep{cw84}.

We set the explosion energy of the 15 and 25 $M_\odot$ SN models
to be $E_{51} = 1$, where $E_{51}$ is the explosion energy in units of
$1 \times 10^{51}$ ergs.
We also calculate the explosion of a HN for the 25 $M_\odot$
star model with $E_{51}=20$.

The location of the mass cut of the SN explosions is determined 
as follows.
For the 15 $M_\odot$ star model, we set the location of the mass cut
at $M_r = 1.71 M_\odot$ ($M_r$ being the mass coordinate of a star)
to obtain the ejected $^{56}$Ni mass of 0.07 $M_\odot$.
In the case of 25 $M_\odot$ SN, we set the mass cut location at
$M_r = 1.92 M_\odot$, so that large (Co,Zn)/Fe ratios are obtained.
This location corresponds to the bottom of the layer where electron fraction
$Y_e$ is close to 0.5 \citep{un05,tu07}.
A large amount of $^{56}$Ni is produced in the region of 
$1.92 M_\odot \le M_r \le 2.2 M_\odot$ (see \S 3.2 for details).
Since the location of the mass cut is deep, this model ejects the $^{56}$Ni 
mass of 0.36 $M_\odot$.
This model corresponds to a SN with the explosion energy of $E_{51} \sim 1$ 
and yielding a large $^{56}$Ni amount ($M$($^{56}$Ni)$\sim 0.3 M_\odot$) 
such as SN 2005bf \citep{to05}.
Note that the mixing-fallback model is not applied to these cases.

In the case of the 25 $M_\odot$ HN model, we apply the mixing-fallback
model \citep{un05,tu07}.
This model approximates aspherical effects of HN explosions \citep{tm07}.
During aspherical explosion, the central core of a star first collapses.
Then, jet-like explosion ejects the surrounding stellar materials.
Some of the materials in an inner region may fallback to the central
core from off-axis direction of the jet.
There are three parameters in the mixing-fallback model; the initial
mass cut $M_{{\rm cut}}$(ini), the outer boundary of the mixing region
$M_{{\rm mix}}$(out), and the ejection factor $f$.
The initial mass cut corresponds to the mass of the initially collapsing
core.
The outer boundary of the mixing region corresponds to the location
where the surrounding materials infall to the core during the explosion.
The ejection fraction is the ejected material fraction in the mixing
region.
Details of the mixing-fallback model have been described in \citet{tu07}.

The values of the mixing-fallback parameters should be related to the 
explosion features such as asphericity and energy deposition rate \citep{tm07}.
However, the explosion mechanism has not been clarified.
Therefore, we determine these parameter values similar to case A 
in \citet{tu07}.
The location of the initial mass cut is set as 
$M_{{\rm cut}}({\rm ini}) = 1.92 M_\odot$, which is the bottom of the 
$Y_e \sim 0.5$ layer.
This condition deduces large Zn/Fe ratio \citep{un02,un05}.
The outer boundary of the mixing region is determined as 
$M_{{\rm mix}}({\rm out}) = 3.83 M_\odot$ where the mass fraction of 
$^{56}$Ni decreases to $10^{-3}$.
The ejection factor is determined as $f = 0.065$ in order to yield 
[O/Fe] = 0.5.
The electron fraction in complete and incomplete Si burning regions is not
modified in this study.
The $^{56}$Ni mass of each explosion model is listed in Table 1.


\subsection{SN Neutrinos and Nucleosynthesis Models}

A huge amount of neutrinos are emitted during SN explosions.
For spherically symmetrical SN explosions, neutrinos are emitted 
isotropically from the neutrino sphere and carry the majority of 
gravitational energy of a proto-neutron star.
On the other hand, aspherical SNe and jet-like explosion would
emit neutrinos anisotropically.
In this case neutrino irradiation may be very strong in a specific direction.
Since we assume spherically symmetrical SN explosions,
we assume isotropic neutrino emission.


The strength of the neutrino flux is determined by the total neutrino
energy.
The adopted values of the total neutrino energy are $E_\nu = 
3 \times 10^{53}$ ergs, $9 \times 10^{53}$ ergs, and $3 \times 10^{54}$ ergs.
The first value corresponds to normal neutrino irradiation and is almost 
equal to the gravitational binding energy of a 1.4 $M_\odot$ 
neutron star \citep{lp01}.
We assume that the neutrino luminosity decrease exponentially with time.
The decay time of the neutrino luminosity is set to be 3.0 s.

The spectra of the SN neutrinos are assumed to obey Fermi distribution
with zero chemical potentials.
The temperatures of $\nu_{\mu,\tau}$, $\bar{\nu}_{\mu,\tau}$ and 
$\nu_{\rm e}$, $\bar{\nu}_{\rm e}$ are set to be $T_\nu = 6$ MeV/$k$ and 
4 MeV/$k$, respectively.
This set of the neutrino temperatures is the same as in \citet{rh02}
and \citet{yu05}.
We note that \citet{ww95} provided the SN yields adopting 
the $\nu_{\mu,\tau}$ and $\bar{\nu}_{\mu,\tau}$ temperature of 
$T_\nu = 8$ MeV and $E_\nu = 3 \times 10^{53}$ ergs.
Therefore, we briefly mention the yields obtained using this parameter set.
Detailed nucleosynthesis during the SN explosions is
calculated by postprocessing using a code of \citet{ht96}
to which the $\nu$-process reactions in \citet{yu05} were added.

We should note that the evolution of neutrino luminosity and energy
spectra of HNe may be different from that of SNe.
The neutrino emission from rapidly accreting disks surrounding black holes
during gamma ray bursts and HNe has been investigated 
\citep[e.g.,][]{ms07}.
They showed that $\nu_e$ and smaller amount of $\bar{\nu}_e$ are mainly
emitted from the accretion disks with slightly hard energy spectra compared
with those from proto-neutron stars.
However, the evolution of neutrino emission during HN explosions
is still largely uncertain.
On the other hand, the main contribution of the $\nu$-process is 
neutral-current interactions, which do not depend on neutrino flavors
\citep{ww95,yt04}.
Furthermore, the yields of the $\nu$-process products depend on both
neutrino energy spectra and the total neutrino energy.
Roughly speaking, a larger total neutrino energy corresponds to harder
neutrino energy spectra \citep{yk05}.
Therefore, we set the total neutrino energy as a parameter.

The value of total neutrino energy $E_\nu = 3 \times 10^{54}$ ergs may be
too large even in the HN model.
If an element is not produced abundantly through the $\nu$-process with
$E_\nu = 3 \times 10^{54}$ ergs, the $\nu$-process contribution should be
very small for the production of the element.

\begin{figure}
\plotone{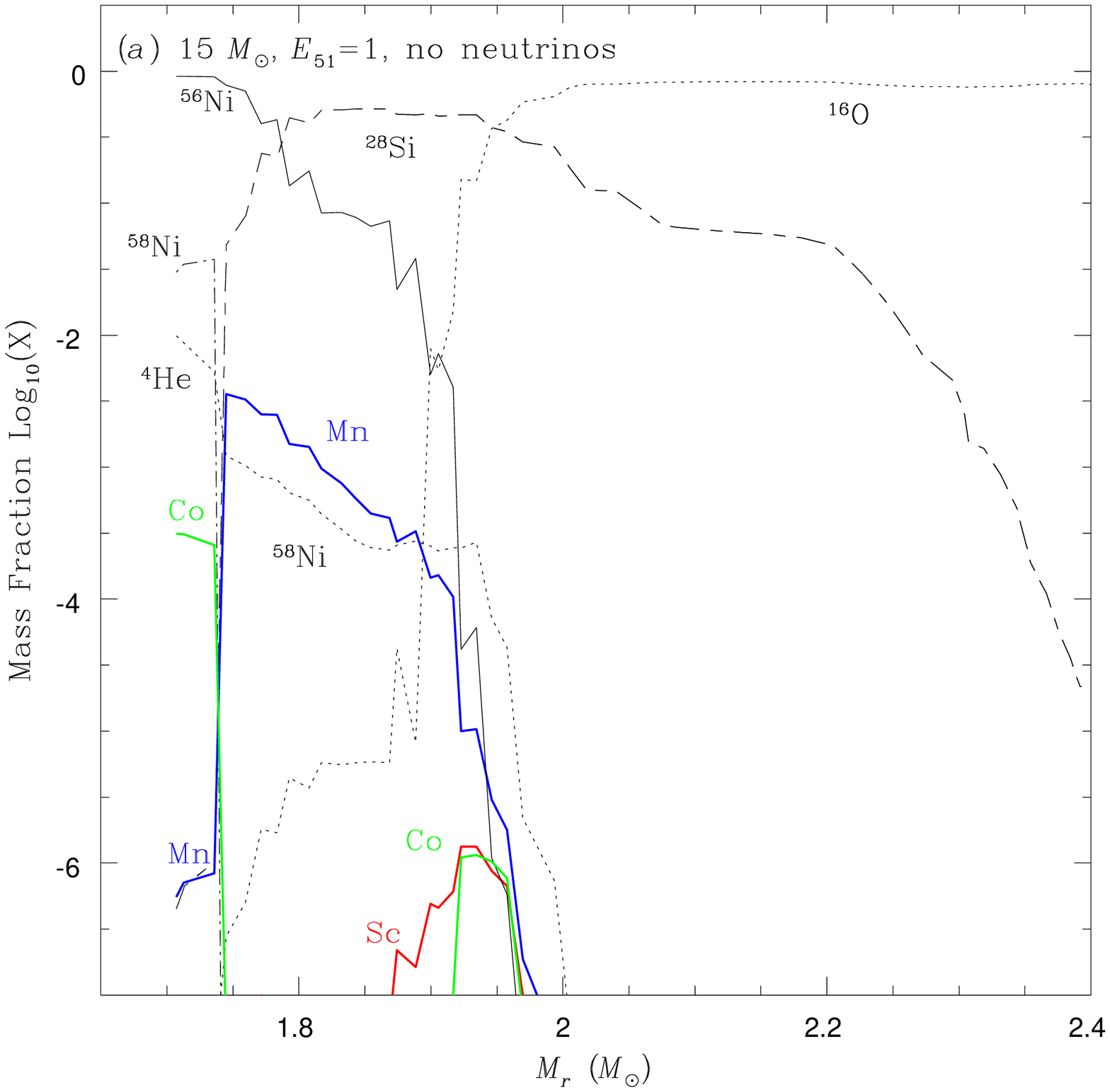}
\plotone{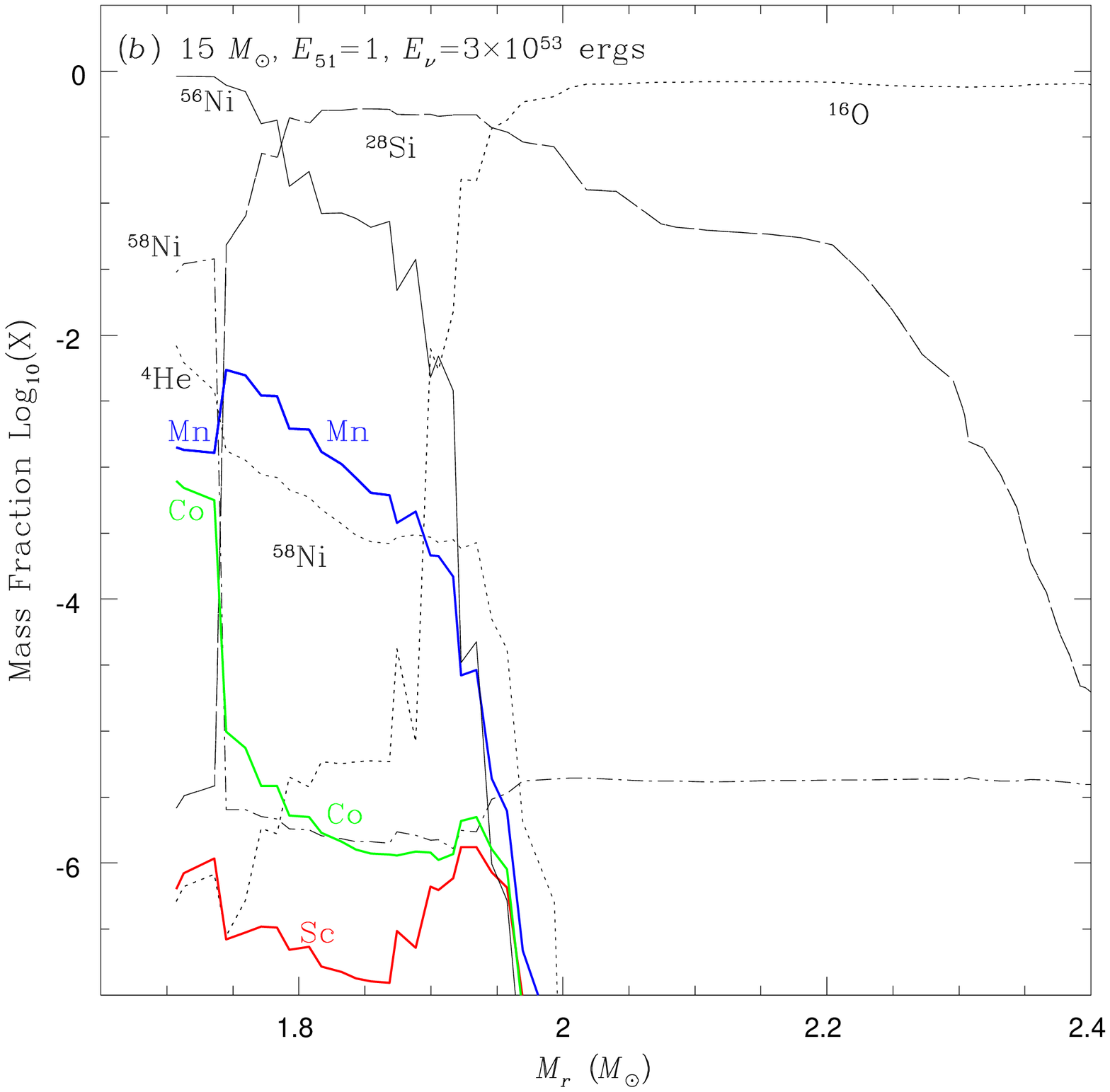}
\plotone{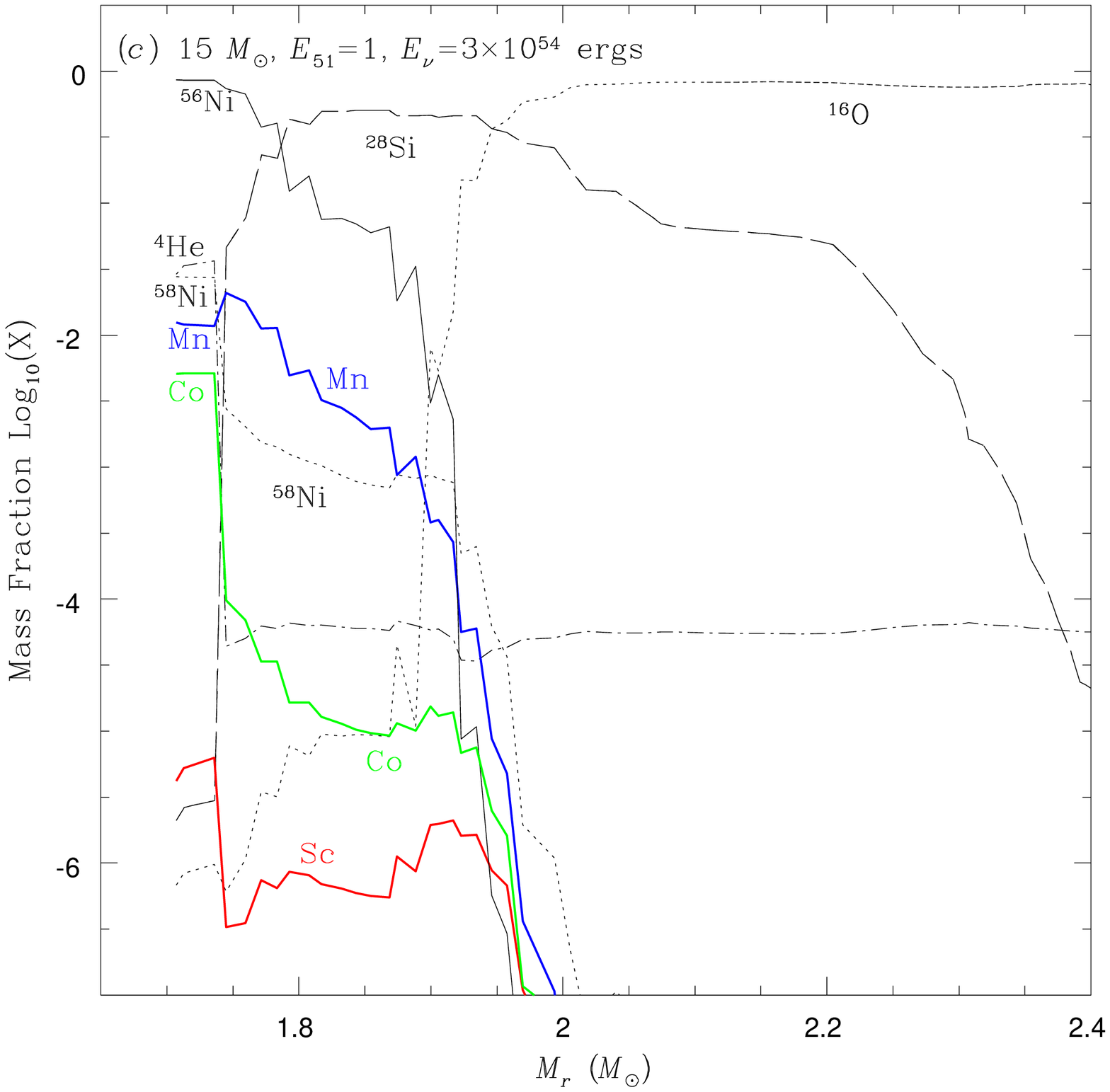}
\caption{
Mass fraction distribution after SN explosions of the 15 $M_\odot$ 
Pop III star model in the cases of ($a$) the $\nu$-process unconsidered,
($b$) the $\nu$-process with $E_{\nu} = 3 \times 10^{53}$ ergs, and 
($c$) the $\nu$-process with $E_{\nu} = 3 \times 10^{54}$ ergs.
Red lines, blue lines, and green lines correspond to the mass fractions
of Sc, Mn, and Co, respectively.
}
\end{figure}

\begin{deluxetable}{ccccc}
\tablecolumns{4}
\tablewidth{0pc}
\tablecaption{The $^{56}$Ni mass and the ratios [Sr/Fe], [Mn/Fe], and [Co/Fe] 
of Pop III SN and HN models.}
\tablehead{
\colhead{$E_\nu$} & \colhead{$M$($^{56}$Ni)} &
\colhead{[Sc/Fe]} & \colhead{[Mn/Fe]}        & \colhead{[Co/Fe]} \\
\colhead{(ergs)}  & \colhead{($M_\odot$)} &
\colhead{} & \colhead{} & \colhead{}
}
\startdata
\cutinhead{15 $M_\odot$ SN}
No neutrinos & $0.072$ & $-1.46$ & $-0.53$ & $-1.10$ \\
$3 \times 10^{53}$ & $0.072$ & $-1.21$ & $-0.29$ & $-0.73$ \\
$9 \times 10^{53}$ & $0.071$ & $-0.99$ & $-0.04$ & $-0.34$ \\
$3 \times 10^{54}$ & $0.068$ & $-0.69$ & $0.36$ & $0.16$ \\
$3 \times 10^{53}$ ($T_{\nu}$=8 MeV) & $0.072$ & $-1.15$ & $-0.20$ & $-0.58$ \\
\cutinhead{25 $M_\odot$ SN} 
No neutrinos & $0.356$ & $-1.73$ & $-0.83$ & $-0.95$ \\
$3 \times 10^{53}$ & $0.353$ & $-1.09$ & $-0.45$ & $-0.42$ \\
$9 \times 10^{53}$ & $0.348$ & $-0.71$ & $-0.12$ & $-0.14$ \\
$3 \times 10^{54}$ & $0.327$ & $-0.29$ & $0.28$ & $0.43$ \\
$3 \times 10^{53}$ ($T_{\nu}$=8 MeV) & $0.353$ & $-0.91$ & $-0.34$ & $-0.28$ \\
\cutinhead{25 $M_\odot$ HN}
No neutrinos & $0.065$ & $-1.13$ & $-1.85$ & $-0.13$ \\
$3 \times 10^{53}$ & $0.065$ & $-1.00$ & $-1.45$ & $-0.19$ \\
$9 \times 10^{53}$ & $0.065$ & $-0.81$ & $-1.12$ & $-0.23$ \\
$3 \times 10^{54}$ & $0.064$ & $-0.44$ & $-0.61$ & $-0.25$ \\
$3 \times 10^{53}$ ($T_{\nu}$=8 MeV) & $0.065$ & $-0.92$ & $-1.32$ & $-0.14$ \\
\enddata

\end{deluxetable}

\section{Abundance Distributions of Sc, Mn, and Co}

\subsection{15 $M_\odot$ Supernova}

We explain the mass fractions of Sc, Mn, and Co as well as 
the contribution of the $\nu$-process for their production.
Figure 1$a$ shows the distributions of the mass fractions of Sc, Mn, and Co 
in the 15 $M_\odot$ SN model without taking into account the 
$\nu$-process.
In this figure, the main product in the range of
$1.71 M_\odot \le M_r \le 1.74 M_\odot$ is $^{56}$Ni.
In this region a large amount of iron-peak elements are produced through 
complete Si burning.
We call this region the Ni layer.
In the region outside the Ni layer, incomplete Si burning occurs during
the explosion and a large amount of Si as well as a smaller amount of 
iron-peak elements are produced.
We call this region ($1.74 M_\odot \le M_r \le 1.95 M_\odot$) the Si/S layer.

The characteristics of the mass fractions and main production sites of
Sc, Mn, and Co without the $\nu$-process are explained as follows.
\begin{enumerate}
\item
Small mass fraction of Sc is produced at
$1.86 M_\odot \la M_r \la 1.95 M_\odot$.
The mass fraction is about $1 \times 10^{-6}$ at the maximum.
Most of Sc is produced through incomplete Si burning.
\item
Most of Mn is produced in the Si/S layer through incomplete Si burning.
The mass fraction at the maximum is about $3 \times 10^{-3}$, which is much
larger than that of Sc.
A part of Mn is also produced in the Ni layer but the mass fraction 
is much smaller.
\item
Co is also produced in the Ni layer.
The mass fraction of Co is about $3 \times 10^{-4}$ throughout this layer.
\end{enumerate}

Hereafter we consider the effect of the $\nu$-process on the production
of Sc, Mn, and Co.
Figures 1$b$ and 1$c$ show the distributions of the mass fractions taking 
into account the $\nu$-process with $E_\nu= 3 \times 10^{53}$ ergs and
$3 \times 10^{54}$ ergs, respectively.

\begin{figure}
\epsscale{1.0}
\plotone{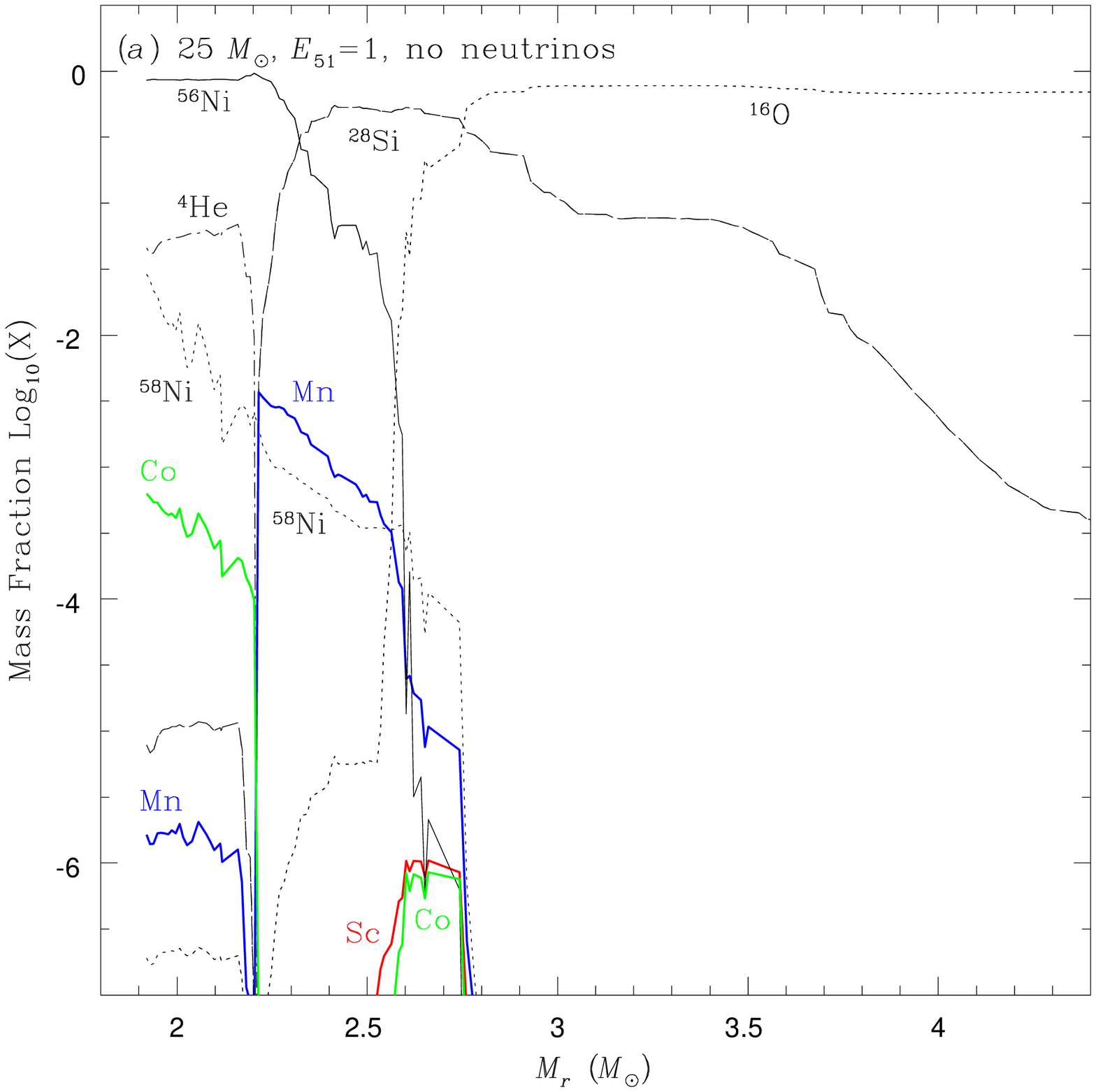}
\plotone{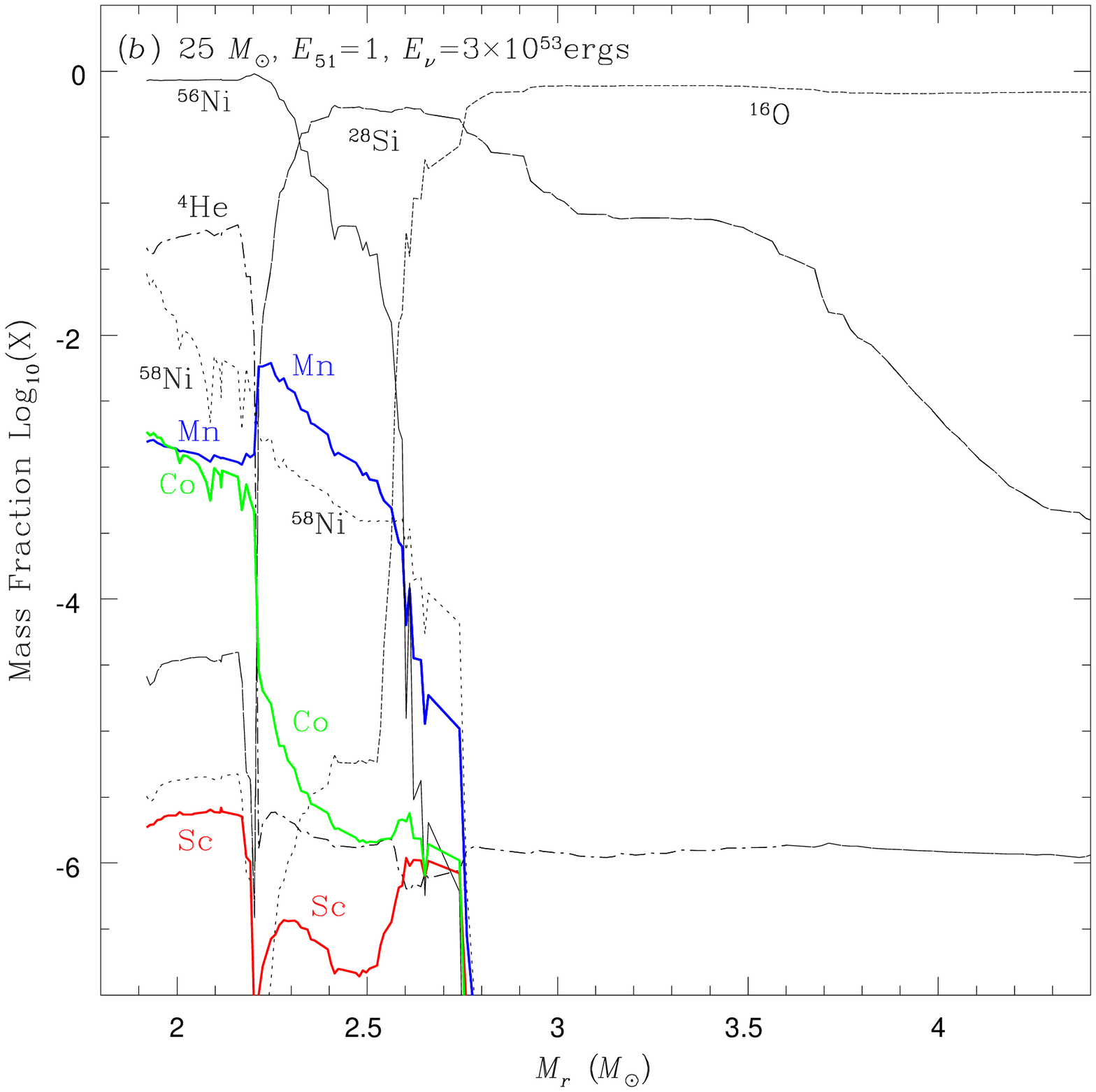}
\plotone{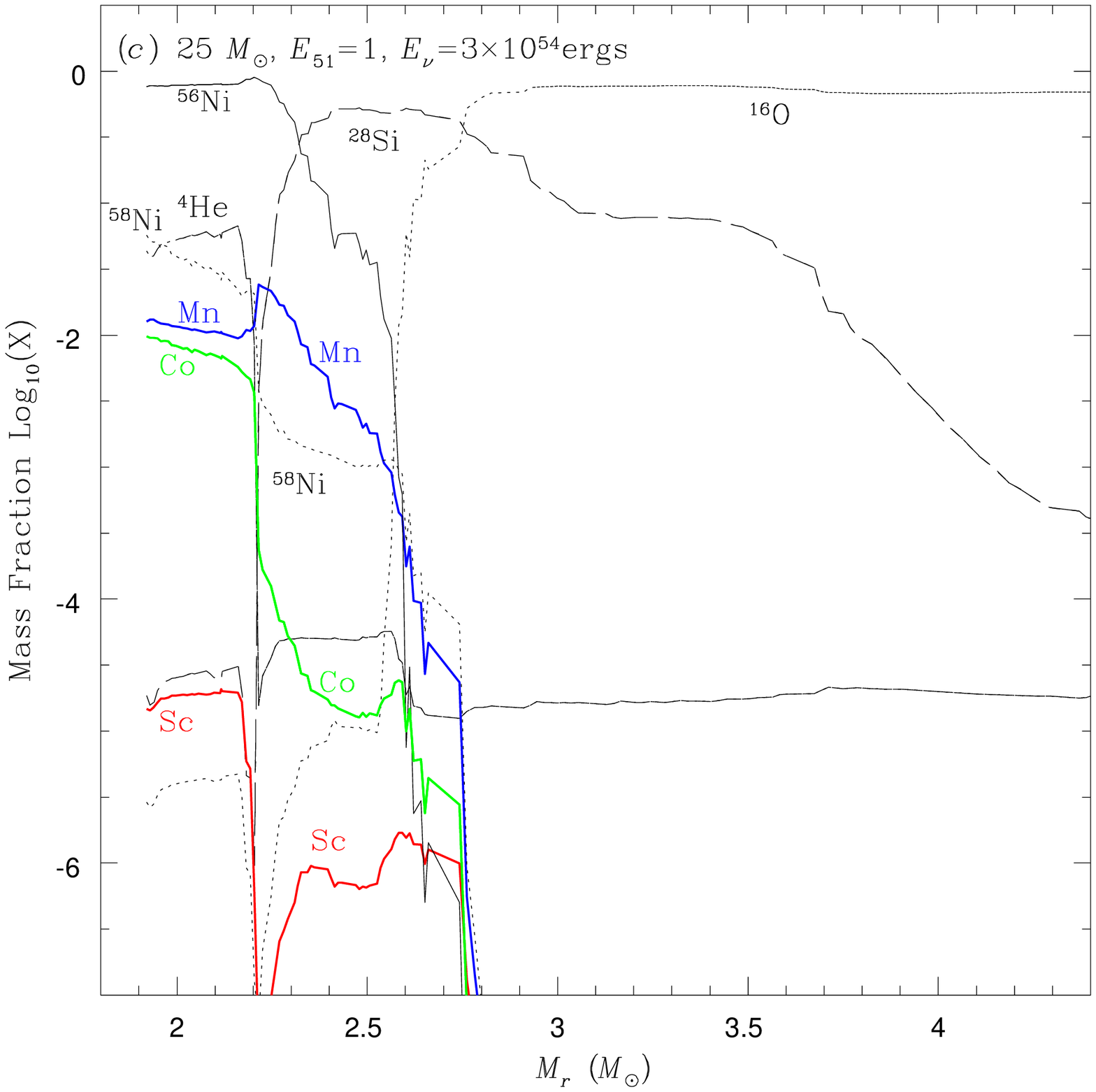}
\caption{
The same as Fig. 1 but the SN explosion of the 25 $M_\odot$ 
Pop III star model.
}
\end{figure}

\subsubsection{Sc}

We see that additional Sc is produced in the Si/S layer 
($1.9 M_\odot \la M_r \la 1.95 M_\odot$) and in the Ni layer.
The mass fraction of Sc is less than $1 \times 10^{-6}$.
Most of the Sc is first produced as $^{45}$Ti and $^{45}$V.
In the Ni layer, a large amount of $^{44}$Ti is produced through complete Si
burning.
At the same time, $^{45}$Ti and $^{45}$V are partly produced through neutron-
and proton-captures of $^{44}$Ti.
Without the $\nu$-process, however, the produced $^{45}$Ti
and $^{45}$V are smoothly captured or decomposed again during the explosion.
With the $\nu$-process, additional protons are produced
through the $\nu$-process, so that some $^{45}$V is produced again
by the proton capture.

\subsubsection{Mn}

We see the increase in the mass fraction of Mn in the Ni layer
by about three orders of magnitude compared with the case without the 
$\nu$-process.
The mass fraction in the Si/S layer also increases by about a factor of 1.4.
During the SN explosion, Mn is produced as $^{55}$Co.
The amount of $^{55}$Mn originally produced in the SN is much smaller.
In the Si/S layer, $^{55}$Co is produced by $^{54}$Fe($p,\gamma)^{55}$Co
or photodisintegration from $^{56}$Ni.
When the $\nu$-process is considered, $^{55}$Co is also produced through
$^{56}$Ni($\nu,\nu'p)^{55}$Co after the production of $^{56}$Ni through
complete and incomplete Si burning.
Since the decay time of the neutrino luminosity is longer than the time
scale of Si burning, the neutrino irradiation proceeds after
ceasing Si burning.

\subsubsection{Co}

When we consider the $\nu$-process, the mass fraction of Co increases
by about a factor of three throughout the Ni layer.
The main production region of Co is not changed when we consider
the $\nu$-process.
Most of Co is produced as $^{59}$Cu and about $10 \%$ of $^{59}$Co is 
produced as $^{59}$Ni.
We see increase in the mass fractions of $^{59}$Cu and $^{59}$Ni similarly
in the Ni layer.
Most of $^{59}$Cu and $^{59}$Ni are produced through the captures of protons
and neutrons by $^{58}$Ni.
The protons and neutrons are produced through the $\nu$-process even
after the cease of complete Si burning.
Therefore, the additional production
raises the mass fraction of Co.

\subsubsection{Larger Neutrino Irradiation}

We also consider the case of larger neutrino irradiation with 
$3 \times 10^{53}$ ergs.
The distributions of the mass fractions are shown in Fig. 1$c$.
We see that the mass fractions of Sc, Mn, and Co in the Ni
layer are larger by a factor of $7 - 9$ than those in the case of
$E_\nu = 3 \times 10^{53}$ ergs.
The dependence of the obtained mass fractions on the total neutrino
energy is slightly smaller than the linear one.
In this case all three elements are mainly produced in the Ni layer
although Mn is mainly produced in the Si/S layer in the case of the former
cases.

\begin{figure}
\epsscale{1.0}
\plotone{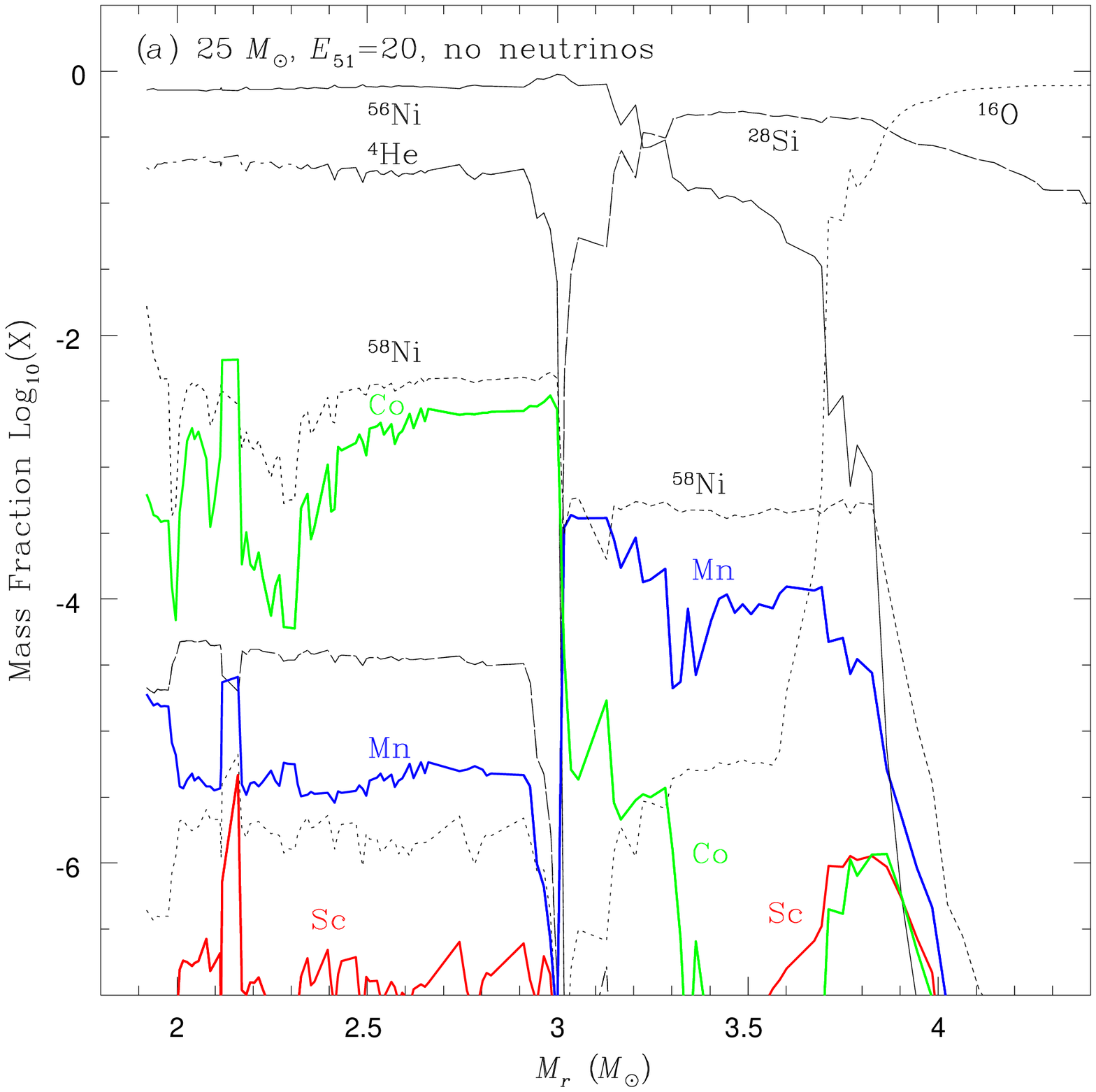}
\plotone{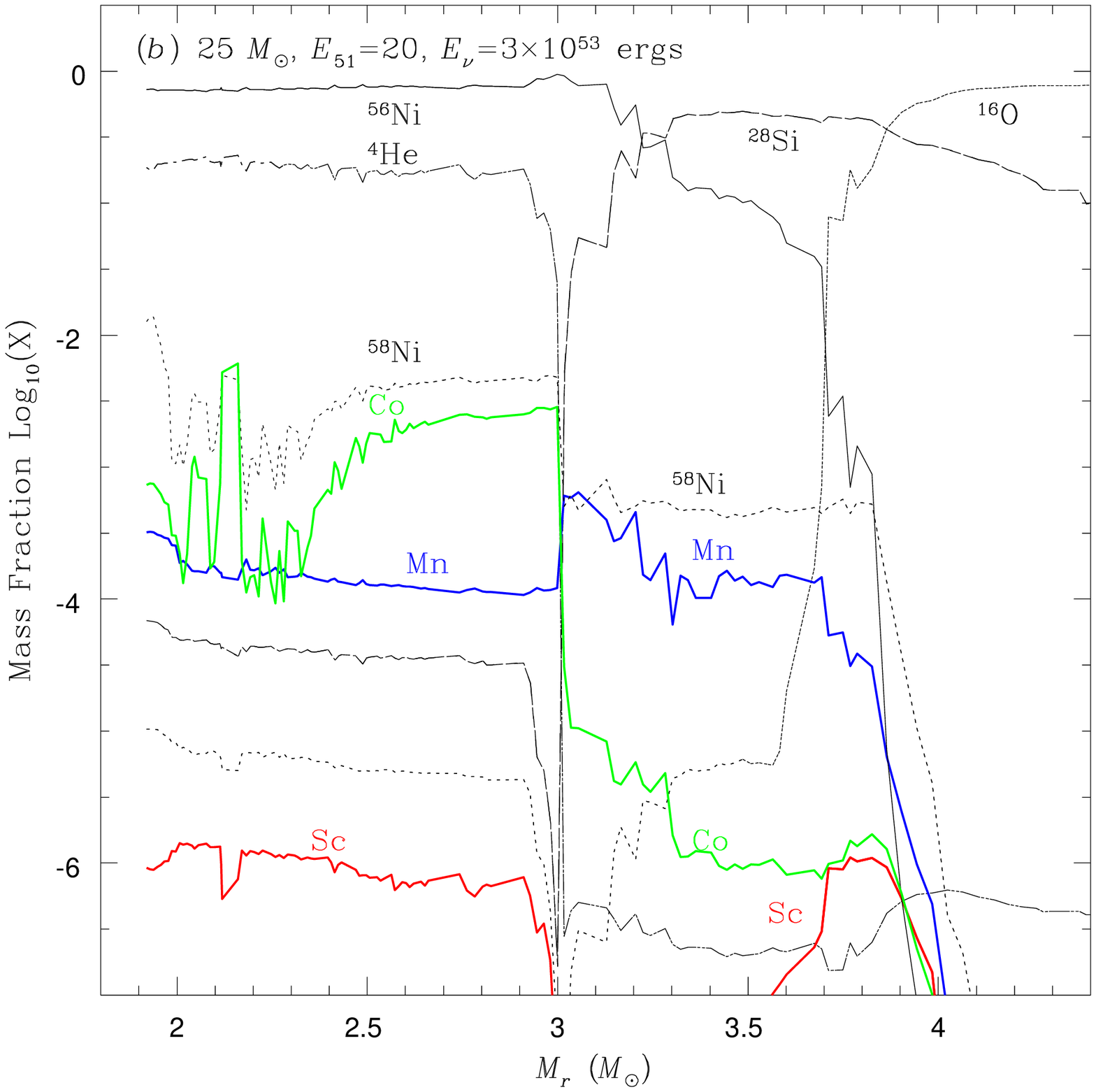}
\plotone{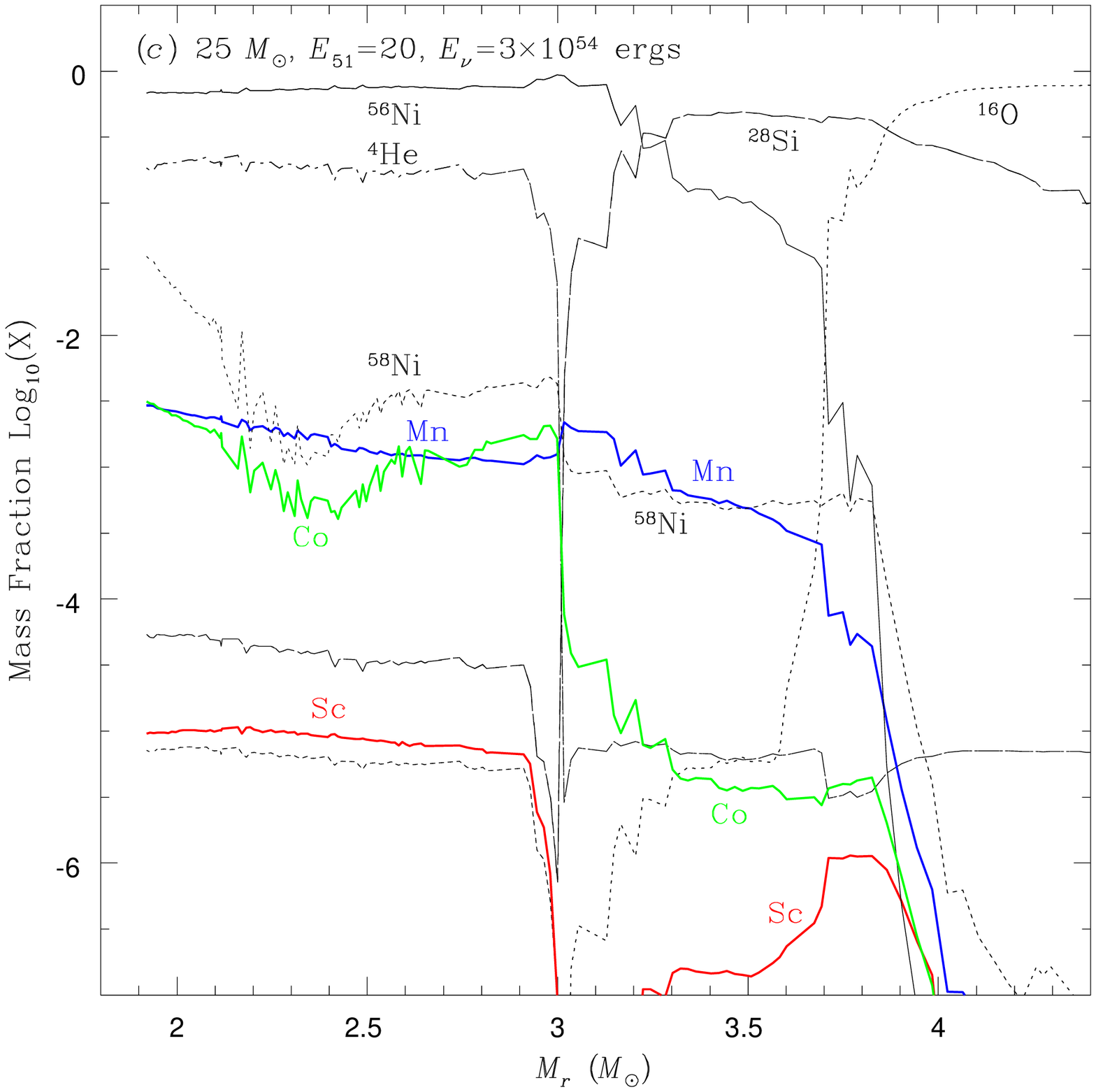}
\caption{
The same as Fig. 1 but the HN explosion ($E_{51}$ = 20) of 
the 25 $M_\odot$ Pop III star model.
}
\end{figure}

\subsection{25 $M_\odot$ Supernova}

The mass fractions of the 25 $M_\odot$ SN model without 
the $\nu$-process is shown in Fig. 2$a$.
In this figure, the Ni layer where complete Si burning occurs is in the range
between $1.92 M_\odot$ and $2.2 M_\odot$ in the mass coordinate.
The Si/S layer which suffered incomplete Si burning is between $2.2 M_\odot$
and $2.75 M_\odot$.
The ejected masses of these layers are larger than the corresponding
layers in the 15 $M_\odot$ SN model.
The distributions of the mass fractions of Sc, Mn, and Co in the 25 $M_\odot$ 
SN model are roughly similar to those of the 15 $M_\odot$ 
SN model.
Most of Sc is produced in outer part of the Si/S layer 
($M_r = 2.6 - 2.75 M_\odot$).
Mn is mainly produced in the Si/S layer, too.
Most of Co is produced in the Ni layer.

The mass fractions of Sc, Mn, and Co with the $\nu$-process of
$E_\nu = 3 \times 10^{53}$ ergs are shown in Fig. 2$b$.
When we consider the $\nu$-process, their increase is seen in the Ni layer.
For Sc, the mass fraction reaches about $2 \times 10^{-6}$.
It is about $1 \times 10^{-3}$ for Mn but the main Mn production is 
still in the Si/S layer.
Increase in the abundance by the $\nu$-process is seen in the 
Si/S layer.
For Co, the abundance increases by about a factor of two.

The mass fractions of Sc, Mn, and Co in the case of a larger neutrino 
irradiation ($E_\nu = 3 \times 10^{54}$ ergs) are shown in Fig. 2$c$.
We obtain larger ones for Sc, Mn, and Co.
The mass fraction of Sc in the Ni layer is about $2 \times 10^{-5}$, which is 
larger than that of the 15 $M_\odot$ SN.
For Mn and Co, the abundances in the Ni layer are close to those of the 
15 $M_\odot$ SN.
The increase in the Mn abundance in the Si/S layer is also seen.

\subsection{25 $M_\odot$ Hypernova}

We explain effects of the $\nu$-process on the production in the
25 $M_\odot$ HN model with $E_{51} = 20$.
Figure 3$a$ shows the distribution of the mass fractions in the 25 $M_\odot$ 
HN model without the $\nu$-process.
The mass coordinate ranges of the Ni and Si/S layers are 
$1.92 M_\odot \le M_r \le 3.0 M_\odot$ and
$3.0 M_\odot \le M_r \le 3.8 M_\odot$, respectively.
The mass contained in each of the two layers for the 25 $M_\odot$ HN
is larger than that of the corresponding layer of the 25 $M_\odot$ 
SN model.
We see that the mass fractions of Sc, Mn, and Co in each
burning region are not so different from those in the 25 $M_\odot$
SN; most of Sc and Mn are produced in the Si/S layer and
Co is produced in the Ni layer.
The average value of the Co mass fraction in the Ni layer is larger
than that in the 25 $M_\odot$ SN.

The distribution of the mass fractions in the 25 $M_\odot$ HN model 
with the $\nu$-process ($E_\nu = 3 \times 10^{53}$ ergs) is shown in Fig. 3$b$.
Increase in the mass fraction by the $\nu$-process is mainly seen for
Sc and Mn.
As shown in the cases of the SNe, additional Sc is 
produced in the Ni layer.
In addition, some Mn is also produced in there.
Since the range of the Ni layer is large, the contribution from the Ni
layer is larger than that from the Si/S layer in this case.
On the other hand, clear effect of the $\nu$-process is not seen for Co.

Figure 3$c$ shows the mass fractions in the 25 $M_\odot$ HN
model with larger neutrino irradiation ($E_\nu = 3 \times 10^{54}$ ergs).
As same as the SN case, the Sc abundance in the Ni layer
is larger by about a factor of ten than that in the case of
the normal neutrino irradiation ($E_\nu = 3 \times 10^{53}$ ergs).
The mass fraction of Mn is similarly larger.
In the Si/S layer, the Mn amount also increases.
On the other hand, the mass fraction of Co does not so change even with
large neutrino irradiation.
Co is produced through complete Si burning in the Ni layer and most of it
is not decomposed during the explosion.
Since HNe produce large amount of Co even without neutrino irradiation
\citep{un05}, the additional production through the $\nu$-process hardly 
affects the Co production.

\begin{figure}
\epsscale{1.0}
\plotone{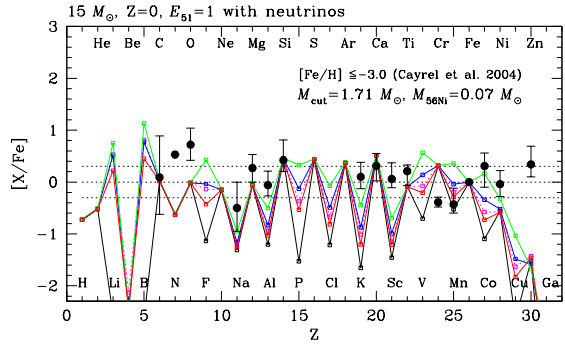}
\caption{
The abundance ratios to Fe in the SN of the 15 $M_\odot$ Pop III star.
The horizontal axis is the atomic number $Z$.
The vertical axis is [X/Fe] (see text).
The black, red, blue, and green lines indicate the abundance ratios
in the cases without the $\nu$-process, with the $\nu$-process and 
$E_{\nu} = 3 \times 10^{53}$ ergs, $9 \times 10^{53}$ ergs, and
$3 \times 10^{54}$ ergs.
The magenta dotted line indicates the one with the $\nu$-process and
$(E_\nu, T_\nu) = (3 \times 10^{53} {\rm ergs}, 8 {\rm MeV})$.
The points indicate the observed abundance ratios averaged in
22 halo stars with [Fe/H] $\le -3.0$ in \citet{ca04}.
Error bars correspond to the abundance ratio ranges of these stars.
}
\end{figure}

\section{Abundance Ratios of Sc, Mn, and Co}

We have shown that some amounts of Sc, Mn, and Co are produced through 
the $\nu$-process and the following capture reactions of neutrons and 
protons. 
This additional production increases the total abundances of these elements
in Pop III SN and HN explosions.
Hereafter we show the distributions of abundance ratios in the 15 $M_\odot$ 
and 25 $M_\odot$ SN and the 25 $M_\odot$ HN models.
We also compare our results with the observed abundances of low mass
EMP stars.
As a reference of the observed abundances, we use the abundance ratios
to Fe averaged in the data of 22 low mass halo stars with [Fe/H] $\le -3.0$ 
in \citet{ca04}.

\subsection{15 $M_\odot$ Supernova}

We show the abundance ratios in the 15 $M_\odot$ SN model 
in Fig. 4.
The ratios of [Sc/Fe], [Mn/Fe], and [Co/Fe] are listed in Table 1.
The abundance ratios of the three elements are much smaller than 
the corresponding solar ratios.
For other elements, the abundance ratios of odd-$Z$ elements are smaller than
those of the even-$Z$ elements in the neighbors.
When we consider the $\nu$-process, 
the ratios [Sc/Fe], [Mn/Fe], and [Co/Fe] increase by $0.24 - 0.37$ dex 
(see Table 1).
Thus, the $\nu$-process raises the abundances of these elements.
Since the Fe abundance is not altered by the 
$\nu$-process, the increase in the abundance ratios is due to the increase 
in the abundances by the $\nu$-process.
Larger irradiation of neutrinos enhances the abundances of these elements.
In the case of $3 \times 10^{54}$ ergs, 
the abundance ratios for Mn and Co are larger than
the corresponding solar ratios (see Table 1 and Fig. 4).
The abundance ratios of V, Mn, and Co are larger than those of the even-$Z$ 
elements Ti, Cr, and Fe, respectively.

We see that [Sc/Fe], [Mn/Fe], and [Co/Fe] for
$E_\nu= 9 \times 10^{53}$ ergs and $3 \times 10^{54}$ ergs increase by 
$0.3-0.4$ dex and $0.7-0.8$ dex, respectively, compared with those with 
$E_\nu = 3 \times 10^{53}$ ergs.
The total neutrino energies with $E_\nu =9 \times 10^{53}$ ergs and 
$3 \times 10^{54}$ ergs are three times and ten times larger than 
that with $E_\nu =3 \times 10^{53}$ ergs.

Simple consideration leads to that the abundances of odd-$Z$ elements produced
through the $\nu$-process are proportional to the neutrino flux, i.e., 
the total neutrino energy.
Numerical simulations of the $\nu$-process show that this property
is approximately correct.
The reaction rates of the $\nu$-process are proportional to the total
neutrino energy.
At the same time, production rates of protons and neutrons produced through
the $\nu$-process are also proportional to the total neutrino energy.
The increase in the production rates brings about the increase in 
the production of odd-$Z$ elements through proton- and neutron-captures.
However, decomposition rates of the odd-$Z$ elements through proton-
and neutron-captures also increase.
Thus, the abundances produced through the $\nu$-process increase less than 
proportional to the total neutrino energy.

We compare the obtained [Sc/Fe], [Mn/Fe], and [Co/Fe] with the corresponding
observed results.
The observed [Sc/Fe] averaged in 22 low mass EMP stars is 0.06.
This value is slightly larger than the solar ratio and is still much
larger than the calculated abundance ratio even in the case of the largest 
neutrino irradiation.
For Mn, the observed [Mn/Fe] is $-0.43$.
This abundance ratio is larger than the ratio without the $\nu$-process and is
reproduced by the result with the $\nu$-process of the normal neutrino
irradiation ($E_\nu = 3 \times 10^{53}$ ergs).
On the other hand, the Mn/Fe ratio in the case of $E_\nu = 3 \times 10^{54}$
ergs is much larger than the largest value in the observed ratios.
The observed [Co/Fe] is 0.31, which is larger than the solar ratio.
It is larger than the value in the SN model even with the largest 
neutrino irradiation.

\begin{figure}
\plotone{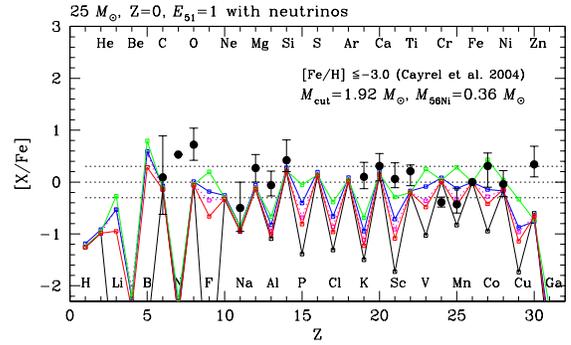}
\caption{
The same as Fig. 4 but in the SN of the 25 $M_\odot$ Pop III star.
}
\end{figure}

\subsection{25 $M_\odot$ Supernova}

Figure 5 shows the abundance ratios of the 25 $M_\odot$ SN model.
When we do not consider the $\nu$-process, the 
abundance ratios of Sc, Mn, and Co are much smaller than the corresponding
solar ratios (see also Table 1).
The $\nu$-process also enhances the production of these elements.
When we consider the $\nu$-process, 
the Sc/Fe ratio in this model is larger than that in the 15 $M_\odot$
SN model.
This is because the seed nucleus of $^{45}$V, i.e., $^{44}$Ti, is abundantly
produced in complete Si burning for the 25 $M_\odot$ SN.
However, the Sc/Fe ratio is smaller than the observed ratio even in the case 
of the largest neutrino irradiation.
On the other hand, the Mn/Fe ratio in this model well reproduces the 
observed ratio in the case with the $\nu$-process of normal
neutrino irradiation.
The model with the largest neutrino irradiation 
($E_\nu = 3 \times 10^{54}$ ergs) overproduces Mn.
The observed Co/Fe ratio is reproduced by the 25 $M_\odot$ SN 
model with the largest neutrino irradiation $E_\nu = 3 \times 10^{54}$ ergs; 
large neutrino irradiation is required similarly to the 15 $M_\odot$ 
SN case.

\begin{figure}
\plotone{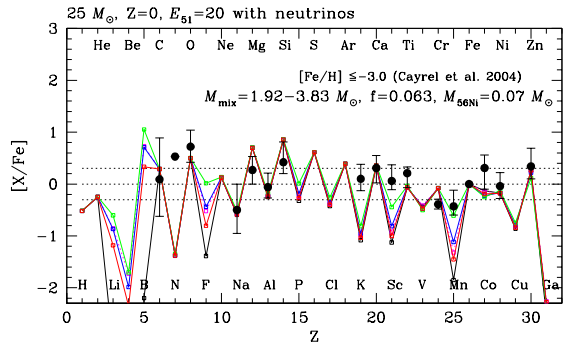}
\caption{
The same as Fig. 4 but in the HN ($E_{51} = 20$) of the 
25 $M_\odot$ Pop III star.
}
\end{figure}

\subsection{25 $M_\odot$ Hypernova}

Figure 6 shows the distribution of the abundance ratios of the 25 $M_\odot$
HN model with $E_{{\rm 51}} = 20$.
In this model we considered mixing-fallback as described in \S 2.1.
When we do not consider the $\nu$-process, 
the [Sc/Fe] and [Mn/Fe] are $-1.1$ and $-1.9$, respectively (see also Table 1).
The Mn/Fe ratio is much smaller than the corresponding
ratios in the other two models.
Mn is mainly produced in the Si/S layer without the $\nu$-process
(see Fig. 3$a$), whereas the range of the Ni layer in the HN is larger
than in the 25 $M_\odot$ SN.
On the other hand, the Co/Fe is larger than
those in the other SN models (see also Table 1).
As shown in Fig. 3$a$, the HN provides a large amount of Co through
complete Si burning.
The mass fraction of Co in the Ni layer is of order $10^{-4}$ for
the 15 $M_\odot$ and 25 $M_\odot$ SN models, and  
$2 - 3 \times 10^{-3}$ for the 25 $M_\odot$ HN.

When we consider the $\nu$-process, the abundance ratios of Sc and Mn increase
with the total neutrino energy.
However, the Co abundance scarcely changes even taking into account the 
$\nu$-process.
The $\nu$-process raises the abundances of Sc and Mn in the Ni layer.
Although it also enhances the Co abundance there, the contribution from the
$\nu$-process is hindered by large Co production through complete Si burning.

We also compare our results with the observed ones.
The obtained Sc/Fe ratio is smaller than the observed ratio.
Even in the case of the largest neutrino irradiation, the abundance ratio is
smaller by about 0.5 dex.
The Mn/Fe ratio is marginally consistent with the observed ratio in the 
largest neutrino irradiation case.
At the same time, the Co/Fe ratio is also marginally consistent with the 
observed ratio despite the $\nu$-process; the smallest Co/Fe ratio is
reproduced.
The HN model reproduces the observed Mn/Fe ratio
when the $\nu$-process with large neutrino irradiation is taken into account.
It reproduces the observed Co/Fe ratio even without the $\nu$-process.

When we do not consider the $\nu$-process, the abundance ratios of 
Sc/Fe, Mn/Fe, and Co/Fe in the Pop III SNe are smaller than 
the corresponding observed ratios.
On the other hand, the observed Mn/Fe ratio is reproduced by the 15 $M_\odot$
and 25 $M_\odot$ SN models with the $\nu$-process.
The observed Co/Fe ratio is marginally reproduced by the 25 $M_\odot$ 
HN model even without neutrino irradiation.
Thus, we can say that the $\nu$-process in the Pop III SNe 
produce most of Mn observed in low mass EMP stars.
The $\nu$-process also produces Sc but the abundance is still short of the 
reproduction of the observed Sc/Fe ratios.
HNe may provide Co being abundant enough to reproduce 
the observed Co/Fe ratios.

\subsection{$\nu$-Process with $T_\nu = 8$ MeV}

We briefly mention the abundance ratios in the case of the $\nu$-process
with $T_\nu = 8$ MeV and $E_\nu = 3 \times 10^{53}$ ergs, which has been
adopted in \citet{ww95}.
The abundance patterns in the 15 and 25 $M_\odot$ SN and the
25 $M_\odot$ HN models are shown as dotted lines in Figs. $4-6$.
The values of [Sc/Fe], [Mn/Fe], and [Co/Fe] are tabulated in Table 1.
The yields of Sc, Mn, and Co in this neutrino parameter set are larger 
by about 0.1 dex than those in the case of $T_\nu = 6$ MeV.
They are still smaller than those in the case of $T_\nu = 6$ MeV and
$E_\nu = 9 \times 10^{53}$ ergs.
Therefore, the dependence on the neutrino temperature is not strong
for the Sc, Mn, and Co yields.

\section{Discussion} 

\subsection{Abundance Pattern of VMP Stars}

The abundance pattern of VMP stars are considered to reflect partly mixed
interstellar materials in the early Universe.
Recently, the averaged abundance pattern of VMP stars with 
$-2.7 < [{\rm Fe/H}] < -2.0$ in \citet{ca04} has been reproduced by the
Pop III SN and HN yields integrated over the Salpeter IMF
function \citep{tu07}.
However, the evaluated [Sc/Fe] and [Mn/Fe] are still short of the observed
ratio.
We discuss the enhancement of these ratios by the $\nu$-process.

\citet{tu07} evaluated the Sc/Fe ratio as [Sc/Fe] = $-1.5$ whereas the 
observed value is [Sc/Fe] = $0.12 ^{+0.24}_{-0.14}$.
In this study, we showed that the $\nu$-process raises the Sc yield by 1.1
dex at the maximum.
However, this enhancement is still short for reproducing the observed value.
Production sites other than the $\nu$-process is needed to reproduce it
(see \S 5.2).

The [Mn/Fe] ratio observed in VMP stars is $-0.48 ^{+0.10}_{-0.24}$.
The evaluated value of [Mn/Fe] is $-1.14$.
When we take into account the $\nu$-process, this value increases by 
$0.24 - 0.4$ dex with $E_\nu = 3 \times 10^{53}$ ergs and $0.49 - 0.73$ dex
with $E_\nu = 9 \times 10^{53}$ ergs.
The increase by the $\nu$-process with $E_\nu = (3-9) \times 10^{53}$ ergs
will provide better fitting of the Pop III SN and HN yields
to [Mn/Fe] in VMP stars.

The Pop III SN and HN models in \citet{tu07} indicate
[Co/Fe] close to the lowest ratio in VMP stars;
[Co/Fe] = $0.29 ^{+0.17}_{-0.27}$.
Taking into account the $\nu$-process with $E_\nu = 3 \times 10^{53}$ ergs,
[Co/Fe] increases by $0.37 - 0.53$ dex in SNe and it does not in
HNe.
This enhancement is enough to reproduce the average [Co/Fe] ratio in VMP stars.
The $\nu$-process of the Pop III SNe and HNe is one of the
preferable sites of the Mn and Co in VMP stars.

\subsection{Other Production Sites of Sc, Mn, and Co}

We have shown that Sc, Mn, and Co are produced through the $\nu$-process
in the Pop III SNe and HNe.
For other than Mn, the $\nu$-process in the SNe and 
HNe does not produce large enough abundances to reproduce 
the abundances observed in low mass EMP stars.
So, we will discuss other synthesis processes proposed in recent studies.

\subsubsection{Sc}

\begin{center}
{\it Aspherical Explosions}
\end{center}

As shown in previous studies \citep[e.g.,][]{un02}, spherically 
symmetrical SN and HN models do not provide enough amount of
Sc to reproduce the observed abundance of low mass EMP stars.
On the other hand, aspherical explosion and jet-like explosion produce
larger amount of Sc \citep{mn03}.
In the case of spherical explosions, low-density structure enables to
supply larger amount of Sc \citep{un05}.
Thus, higher temperature and low density, which are realized by aspherical
explosions, are one of the favorable environments to produce Sc.
The amount of Sc produced in SNe and HNe is also sensitive 
to $Y_e$ in complete Si burning region. 
Large $Y_e$ value ($Y_e > 0.5$) brings about a large amount of Sc \citep{iu06}.

\begin{center}
{\it Nucleosynthesis in Innermost Region of SNe}
\end{center}

Detailed thermal evolution of the innermost region of SN ejecta, 
i.e., the neighbor of the ^^ ^^ mass cut'' has not been solved strictly.
However, recent progress of hydrodynamical calculations has gradually 
revealed thermal evolution in such a deep region.
At the same time, detailed nucleosynthesis in such a deep region has also
been investigated.
\citet{pw05,ph06} investigated the nucleosynthesis in convective bubbles 
in the innermost region of SN ejecta and in early stage of 
neutrino-driven winds.
They used results of two-dimensional hydrodynamical calculations 
by \citet{jb03}.
\citet{fr06a} examined the simulations of one-dimensional
self-consistent treatment of core-collapse SNe with modified neutrino
cross sections to explode.
They also investigated the nucleosynthesis in the innermost region just 
above the mass cut produced self-consistently.

Both of the studies showed that the electron fraction $Y_e$ exceeds 0.5 
in such regions, which enhances the production of  Sc.
In the latter study, they took into account all charged-current 
weak-interactions that changes $Y_e$ in the ejecta.
They obtained that the amount of Sc with $10^{-6} M_\odot$ is produced in the 
innermost region with $Y_e > 0.5$.
The Sc/Fe ratio evaluated in their study is slightly smaller than the 
averaged abundances observed in low mass EMP stars \citep{ca04}.
Thus, the Sc/Fe ratio is larger than our result.
They also showed that a significant amount of nuclei with mass of $A > 64$
in this region, especially light $p$-nuclei such as $^{92,94}$Mo and
$^{96,98}$Ru \citep[the $\nu p$-process;][]{fr06b}.

It is noted that the region of $Y_e > 0.5$ is convectively unstable so that
the region will mix with outer smaller $Y_e$ region in dynamical time scale.
However, they expected that the $Y_e$ remains high in an average sense.
Indeed, the former study took into account the convection in two-dimensional
simulations and obtained $Y_e > 0.5$.
We also note that neutral-current $\nu$-process reactions were not included
in \citet{fr06a,fr06b}.
They also pointed out that neutrino-induced spallation reactions can change
the final abundances of some nuclei.
We expect that these reactions enhance the abundance of Sc and that the Sc/Fe
ratio becomes close to the observed ratio.

It is also noted that materials blown off in neutrino-driven winds 
in early time after a SN explosion can achieve large $Y_e$ value.
Accordingly, the $rp$-process can take place in the materials.
This synthetic process is strongly preferable for the production of 
the light $p$-nuclei \citep{wa06}.
However, the contribution of the $rp$-process in neutrino-driven winds 
to Sc production is small.

\subsubsection{Mn}


The amount of Mn produced in SNe and HNe is strongly sensitive
to electron fraction $Y_e$ in incomplete Si burning region \citep{un02,un05}.
Their HN models modifying $Y_e$ value to $Y_e \simeq 0.4995 - 0.4997$ 
produce enough amount of Mn to reproduce the observed one in low mass EMP 
stars.
However, HN models, whose $Y_e$ in incomplete Si burning region is
not modified, indicated that the amount of Mn is less than the observed value.
Aspherical explosions suppress Mn production \citep{mn03}.
We have shown that the Mn production during SN explosions is mainly due to
the $\nu$-process rather than to the incomplete Si burning.
The $\nu$-process with normal neutrino irradiation ($E_\nu = 3 \times 10^{53}$
ergs) in SNe reproduces the Mn/Fe ratio averaged in low mass EMP stars
even without $Y_e$ modifications.
In the case of $E_\nu = 9 \times 10^{53}$ ergs, the Mn amount marginally
reproduces the upper limit of the observed Mn abundance.
In the case of $E_\nu = 3 \times 10^{54}$ ergs, however, the SN 
models overproduce Mn.
Thus, the $\nu$-process in the Pop III SNe is one of 
the main production processes of Mn observed in low mass EMP stars.
The observed abundance constrains the total neutrino energy $E_\nu$ in 
SNe.
The observational constraint is $E_\nu \la 3 - 9 \times 10^{53}$ ergs.

\subsubsection{Co}

A large amount of Co can be produced through complete Si burning in aspherical
explosion \citep{mn03}.
For spherical explosions, \citet{un05} pointed out that the Co
abundance is significantly enhanced for $Y_e > 0.5$ in the case of HNe.
However, the Co/Fe ratio in \citet{fr06a} is still smaller than
the observed ratio in \citet{ca04}.
Their model corresponds to a normal SN explosion.
Therefore, HNe ($E_{51} > 10$) 
with $Y_e \ga 0.5$ produce Co of which amount reproduces 
observed Co/Fe ratio in low mass EMP stars.
Neutral-current $\nu$-process reactions may help additional production in
the deep region just above the mass cut.
However, in the case of HNe, large amount of protons are
also produced through $\alpha$-rich freeze out.
This effect may hinder the proton production by the $\nu$-process.

\subsection{Comparison with Other SN Nucleosynthesis Models}

The nucleosynthesis in Pop III SNe have been conducted
in several groups.
Here, we briefly compare with the abundances of Sc, Mn, and Co of
15 $M_\odot$ and 25 $M_\odot$ Pop III SN models in \citet{ww95},
hereafter abbreviated to WW95, (Z15A and Z25B models) and in \citet{cl04}, 
abbreviated to CL04. 
The SN nucleosynthesis models in WW95 include the $\nu$-process.
The temperature of $\nu_{\mu,\tau}$ in WW95 is assumed to be 8 MeV, which
is larger than the one adopted in this study.
The SN nucleosynthesis with updated physics inputs of WW95 has been
shown in \citet{rh02} but the Pop III SN nucleosynthesis has not
been calculated.
The nucleosynthesis models in CL04 do not include the $\nu$-process.
The explosion models in CL04 were set to be the ejected mass of $^{56}$Ni 
equal to 0.1 $M_\odot$.

\citet{ww95} indicate [Sc/Fe]=$-0.23$ for $15 M_\odot$ model; the Sc/Fe
is much larger than our result.
On the other hand, CL04 indicates Sc/Fe slightly larger than our results
of SN models without the $\nu$-process ([Sc/Fe]=$-1.58$ for
25 $M_\odot$ model). 
Although the Sc/Fe in WW95 is much larger than the other models, it is
still smaller than the observed ratio.
Therefore, other synthesis processes such as discussed above are necessary
for Sc production.

The Mn/Fe shown in WW95 is smaller than the observed ratio
([Mn/Fe]$\le -0.50$).
The 15 $M_\odot$ model in CL04 also indicates the Mn/Fe smaller
than the observed one ([Mn/Fe]$=-0.56$).
The 25 $M_\odot$ model in CL04 indicates larger Mn/Fe ratio, which is 
comparable to the observed one ([Mn/Fe]$=-0.33$).
We showed that the $\nu$-process increases the Mn abundance, but difference
of stellar evolution models may affect the abundance.

The 15 $M_\odot$ models of WW95 and CL04 show [Co/Fe] of $-0.22$ and $-0.11$,
which are larger than the ratio of our SN models. 
Their 25 $M_\odot$ models indicate the Co/Fe similar to our results
without the $\nu$-process.
The evaluated Co/Fe ratios are still short of the observed ratio, so that
HNe will have an important role for Co production in the early Universe.

\subsection{The $\nu$-Process for Other Elements}

\subsubsection{Li and B}

Among Li, Be, and B, $^7$Li and $^{11}$B are mainly produced through the
$\nu$-process.
When the $\nu$-process is not considered, the Li/Fe and B/Fe ratios are
much smaller than the corresponding solar ratio.
SN explosions with the neutrinos of $E_{\nu} = 3 \times 10^{53}$ ergs
bring about Li yield ratio of [Li/Fe] $\ga -1$.
They also produce B of which yield ratio to Fe is larger than the solar
ratio.
Therefore, SN explosions may contribute to the B production in
the early universe.
The $^7$Li is mainly produced as $^7$Be in the He-rich region.
The $^{11}$B is produced as $^{11}$C and $^{11}$B in C-enriched oxygen layer
and C-enriched He layer.
The Li and B production in HNe is less effective than that in 
SNe.
Higher maximum temperature in the He layer decomposes $^7$Be and $^{11}$C
produced through the $\nu$-process and the following $\alpha$-captures.
In addition, strong explosion makes the region of C-rich oxygen layer small.

\subsubsection{F}

The $\nu$-process is a main F production process in SNe \citep{wh90}.
The F/Fe ratio including the $\nu$-process is
much larger than the one without the $\nu$-process.
The seed nucleus of F in the $\nu$-process is $^{20}$Ne.
Most of F is produced in O and Ne-enriched region.
Stellar mass dependence of the F/Fe ratio is small in the SN models.
On the other hand, the HN model shows smaller F/Fe ratio.
The HN explosion brings about stronger explosive Ne burning, 
so that the O and Ne-enriched region becomes smaller.

\subsubsection{Na and Al}

Figures $4-6$ show that the $\nu$-process scarcely affects the yields of Na 
and Al.
Furthermore, the yield ratios are much smaller than the observed ratios
even in the case of the largest neutrino irradiation.
Although the $\nu$-process enhances the abundances of Na and Al in incomplete 
Si burning region, Na and Al are mainly produced in carbon and neon shells.
The yields produced in the outer two shells strongly depend on the 
nucleosynthesis in stellar evolution.
\citet{iu05} successfully reproduced relatively large [Na/Fe] and [Al/Fe] 
observed in two hyper-metal-poor stars.
They considered that the effect of overshooting during preSN
evolution and extensive matter mixing and fallback during SN
explosions with small explosion energies.

\subsubsection{V}

Vanadium is mainly produced as $^{51}$Mn in incomplete Si burning region
in SNe.
In the case of the HN model, it is mainly produced in complete Si 
burning region.
Additional production through the $\nu$-process from $^{52}$Fe in incomplete
Si burning region contributes to the enhancement of V in SNe.
Low metallicity stars with $-3 \la {\rm [Fe/H]} \la -2$ indicate the
V abundance ratios of $-0.2 \la {\rm [V/Fe]} \la 0.6$ 
\citep[e.g.,][]{ha04,kn06}.
The average abundance ratio is reproduced by the 15 and 25 $M_\odot$
SN models with the $\nu$-process of $E_\nu \sim 9 \times 10^{53}$ ergs.
In order to reproduce the V/Fe ratios, strong neutrino irradiation
in the SNe is favored.
In the case of the HN model, V is also produced through 
the $\nu$-process.
However, the produced abundance is much smaller than that produced in 
complete Si burning region.

\section{Summary}

The chemical compositions of low mass EMP stars are expected to be injected
from one or a few SN or HN explosions evolved from 
Population III massive stars.
The abundance ratios of odd-$Z$ elements to Fe observed in these
stars are close to the corresponding solar ratios.
On the other hand, the abundances produced in complete and incomplete Si 
burning in SN explosion models are still in short.

In this study, we investigated the $\nu$-process of odd-$Z$ iron-peak 
elements, Sc, Mn, and Co, in the SN and HN explosions of 
15 $M_\odot$ and 25 $M_\odot$ Pop III stars.
Then, we compared the abundance ratios of these elements in the SN
and HN models with those observed in the EMP stars.
The obtained results are summarized as follows.
\begin{enumerate}
\item
Sc, Mn, and Co are produced through the $\nu$-process and the following
capture reactions of protons and neutrons in complete and incomplete Si 
burning region of Pop III SNe.
The produced amounts of these elements are roughly proportional to the
total neutrino energy.
\item
The observed Mn/Fe ratio averaged in low mass EMP stars is reproduced
by the 15 $M_\odot$ and 25 $M_\odot$ Pop III SN models 
with the $\nu$-process.
Therefore, the $\nu$-process in Pop III SNe is one of the main synthesis
processes of Mn observed in low mass EMP stars.
The observed Mn abundance constrains the total neutrino energy to 
$E_\nu \la 3 - 9 \times 10^{53}$ ergs.
This $E_\nu$ value is roughly consistent with the gravitational binding
energy of a typical neutron star.
In the case of the 25 $M_\odot$ Pop III HN model, large neutrino 
irradiation enables to reproduce the observed Mn/Fe ratio.
\item
The observed Co/Fe ratio is reproduced by the SN models.
However, required total neutrino energy is $E_\nu \sim 9 - 30$ ergs,
which is larger than typical value of the gravitational binding energy of 
a proto-neutron star.
Complete Si burning in Pop III HNe should be favorable to
Co production rather than the $\nu$-process of SNe
and HNe.
\item
Although larger amount of Sc is produced through the $\nu$-process, the Sc/Fe
ratio is still smaller than the ratio observed in low mass EMP stars.
\item
Explosive nucleosynthesis during aspherical explosions and in a deep region 
just above the mass cut would be favorable to produce Sc and Co.
The $\nu$-process and the capture reactions of protons and neutrons
may also enhance the production of Sc, Mn, and Co in these environments.
\end{enumerate}



\acknowledgments

We would like to thank anonymous referee for valuable comments.
Numerical computations were in part carried out on general common use
computer system at Astronomical Data Analysis Center (ADAC) of National
Astronomical Observatory of Japan.
This work has been supported in part by the Ministry of Education, Culture,
Sports, Science and Technology, Grants-in-Aid for Young Scientist (B)
(17740130), and Scientific Research (S) (18104003), (C) (16540229, 18540231).
T.Y. was supported by the 21st Century COE Program ^^ ^^ Exploring New Science
by Bridging Particle-Matter Hierarchy'' in Graduate School of Science, Tohoku
University.

\end{document}